\documentclass[%
 reprint,%
 amssymb, amsmath,%
 prc,cha,%
superscriptaddress
]{revtex4-2}
\usepackage[colorlinks=true,linkcolor=blue]{hyperref}%
\usepackage{physics}
\usepackage{graphicx}
\usepackage{lipsum}
\usepackage{mwe}
\usepackage{caption}

\usepackage{dcolumn}
\usepackage{bm}
\usepackage{adjustbox}
\usepackage{rotating}
\usepackage{xcolor}
\usepackage{ulem}

\begin{document}

\preprint{APS/123-QED}

\title{Electric dipole transitions in the relativistic quasiparticle random phase approximation at finite temperature}%

\author{Amandeep Kaur}\email[]{akaur.phy@pmf.hr}
\affiliation{Department of Physics, Faculty of Science, University of Zagreb, Bijeni\v{c}ka c. 32,  10000 Zagreb, Croatia}

\author{Esra Y{\"{u}}ksel}\email[]{e.yuksel@surrey.ac.uk}
\affiliation{Department of Physics, University of Surrey, Guildford, Surrey GU2 7XH, United Kingdom}

\author{Nils Paar}\email[]{npaar@phy.hr}
\affiliation{Department of Physics, Faculty of Science, University of Zagreb, Bijeni\v{c}ka c. 32,  10000 Zagreb, Croatia}

\date{\today}

\begin{abstract}
Finite temperature results in various effects on the properties of nuclear structure and excitations of relevance for nuclear processes in hot stellar environments. Here we introduce the self-consistent finite temperature relativistic quasiparticle random phase approximation (FT-RQRPA) based on relativistic energy density functional with point coupling interaction for describing the temperature effects in electric dipole (E1) transitions. We perform a study of E1 excitations in the temperature range $T=$ 0-2 MeV for the selected closed- and open-shell nuclei ranging from $^{40}$Ca to $^{60}$Ca and $^{100}$Sn to $^{140}$Sn by including both thermal and pairing effects. The isovector giant dipole resonance strength is slightly modified for the considered range of temperature, while new low-energy peaks emerge for $E<$12 MeV with non-negligible strength in neutron-rich nuclei at high temperatures. The analysis of relevant two-quasiparticle configurations discloses how new excitation channels open due to thermal unblocking of states at finite temperature. The study also examines the isospin and temperature dependence of electric dipole polarizability ($\alpha_D$), resulting in systematic increase in the values of $\alpha_D$ with increasing temperature, with a more pronounced effect observed in neutron-rich nuclei. The FT-RQRPA introduced in this work will open perspectives for microscopic calculation of $\gamma$-ray strength functions at finite temperatures relevant for nuclear reaction studies. 

\end{abstract}
\maketitle

\section{\label{sec:Intro}Introduction}
The electromagnetic excitations in nuclei have attracted considerable attention, both for their pivotal role in advancing our comprehension of nuclear structure and their contributions to $r$-process nucleosynthesis within stellar environments. \cite{ARNOULD200797,Larsen82,Goriely706,Xu86}. 
Among the various excitation modes in nuclei, the Giant Dipole Resonance (GDR) is one of the most well-known excitation modes and has been extensively studied over the years, both experimentally \cite{Bracco106,CAPOTE20093107,KAWANO2020109} and theoretically \cite{Paar_2007, paar67, Goriely706, Daoutidis86}. The GDR is located at high energies above the neutron separation energy and originates from the collective oscillation of protons against neutrons \cite{Harakeh1, J_Speth_1981}. 

In neutron-rich nuclei, the emergence of a low-energy (pygmy) dipole strength, has also been predicted in many studies (see Refs. \cite{Paar_2007, Bracco106,SAVRAN2013210}). The low-energy dipole strength is interpreted as the oscillations of the neutron skin (neutrons in outer orbitals) against the isospin-symmetric core of the nucleus and arises mainly due to transitions involving neutrons from outer orbitals \cite{Vretenar692, BRACCO2019360, LANZA2023104006}. Although the nature of pygmy dipole strength is currently under discussion, it exhibits a direct correlation with the neutron-skin thickness, electric dipole polarizability, and nuclear matter symmetry energy \cite{Roca-Maza92,Piekarewicz85,Paar103,Bracco106}. Furthermore, the low-energy dipole transitions in the vicinity of the neutron threshold are essential in the calculations of astrophysical reaction rates and cross-sections of photonuclear and radiative-capture reactions \cite{Goriely706,Bracco106,Tonchev773,Larsen107,Daoutidis86}.

Numerous theoretical and experimental investigations have been conducted to explore the isotopic sensitivity of both natural parity electric dipole (E1) and unnatural parity magnetic dipole (M1) responses. These studies revealed a pronounced dependence of dipole strength distributions on the neutron-to-proton (N/Z) ratio
\cite{Paar606,Guliyev47,Kruzic103,Oishi47,Kemah59,Bassauer102,Enders486,Quliyev1014,Tonchev104,Hartmann85}. Given that nuclei can also exist in extreme conditions within stellar environments, there has been a growing interest in understanding the temperature dependence of nuclear excitations. Temperature ($T$) serves as an external probe capable of significantly influencing the structure and dynamics of nuclei \cite{Ravlic2023, Yuksel50,Yuksel96,Litvinova88,Litvinova121}. The experimental data on the GDR at finite temperature is mainly based on the study of the decay from fusion-evaporation reactions, which allow the production of self-conjugate compound nuclei (CN) at high excitation energy \cite{PhysRevLett.97.012501, PhysRevC.84.041304}. Despite being challenging experimentally, numerous studies have been carried out to investigate the $T$-dependence of the dipole response, especially in the GDR region of highly excited nuclei \cite{Bracco22,Heckman555,Santonocito,Wieland97,Yoshida245,Chakrabarty52,Mondal784,PhysRevLett.97.012501, PhysRevC.84.041304}. Recent studies have shown that the GDR width has increased rapidly with temperature between 1 and 3 MeV; however, it gets saturated at much higher temperatures, which results in the gradual disappearance of giant dipole resonance due to excessive broadening \cite{Yoshida245,Chakrabarty52}. On the other hand, experimental analysis of $T$-dependent pygmy dipole strength is still missing. The main complication at the experimental front is to differentiate between the E1 and M1 responses in the low-energy region since the M1 strength is highly fragmented at pygmy dipole excitation energies \cite{Harakeh1}. These E1 and M1 excitation channels can be discriminated on the basis of spin and parity of involved states and can be explored theoretically. Recent developments open perspectives for the on-going first experimental study of pygmy dipole strength in Ni isotopes at temperatures up to $\approx$ 2 MeV \cite{Wieland2023}.

Developing a theory to study the complex structure of dipole strengths at finite temperature (FT) is an ambiguous task. Several extensions of random phase approximation (RPA) were developed to explore the electric multipole nuclear excitations at zero and higher temperatures. A self-consistent finite temperature relativistic RPA (FT-RRPA) based on meson-exchange interaction was successfully employed to study the evolution of isoscalar monopole and isovector dipole modes with temperature \cite{Niu681}. Based on Skyrme-type energy density functional, the finite temperature continuum quasiparticle RPA \cite{Khan731,Litvinova88} and the finite temperature quasiparticle RPA (FT-QRPA) \cite{Yuksel96,Yuksel55} were established to elucidate the nuclear response of open-shell hot nuclei. 
Within the {\it ab initio} approach, the electromagnetic response functions have been studied at finite temperatures for medium-mass nuclei \cite{beaujeault2023zero}.
By adopting a time blocking technique to the Matsubara temperature Green's function formalism, finite temperature relativistic time blocking approximation (FT-RTBA) approach was formulated to study the dipole spectra in the excited nuclei \cite{Litvinova121,Litvinova55}, whereas the study is limited to closed shell nuclei. All of these studies have revealed the emergence of new excited states, particularly in the low-energy region, attributed to the thermal unblocking effect of temperature on single-particle orbitals near the Fermi level. Note that, in open shell nuclei, it is required to include both the finite temperature and pairing effects when considering nuclear excitations at temperatures below the pairing collapse. Considering that previous FT-RTBA study in the relativistic framework is based on meson-exchange interaction with non-linear self-interaction terms, that overestimates the parameters of the symmetry energy, a thorough analysis is required to elucidate the evolution of E1 dipole response at finite temperature, by using relativistic functionals with density-dependent vertex functions providing realistic description of the symmetry energy, that is relevant for studies of isovector dipole response. To address both closed- and open-shell nuclei at finite temperatures, development of a fully self-consistent finite temperature relativistic quasiparticle RPA is required (FT-RQRPA), that also includes pairing correlations both in the description of the ground state and excitations.

In the present work, the finite-temperature extension of fully self-consistent relativistic QRPA (FT-RQRPA) is established for the study of E1 excitations. This framework is based on the relativistic nuclear energy density functional (RNEDF), which includes both pairing and temperature effects. We note that in the relativistic framework, proton-neutron FT-RQRPA has already been introduced and used in studies of charge-exchange transitions, such as Gamow-Teller or higher order forbidden transitions \cite{PhysRevC.101.044305, Ravlic102,Ravlic104}. However, transitions without charge-exchange require formalism with different residual interaction channels. Following the development of the FT-RQRPA in the present work, the objective is to explore the temperature evolution of isovector E1 (J$^\pi$=1$^-$) excitations in the high-energy as well as low-energy regions for $^{40-60}$Ca and $^{100-140}$Sn isotopic chains, and explore in detail their structure.

The paper is organized as follows: Sec. \ref{sec:Formalism} provides an overview of the FT-RQRPA theoretical framework. Subsequently, the details of FT-RQRPA calculations are discussed in Sec. \ref{sec:results}. The temperature dependence of isovector E1 strength distributions in Ca and Sn isotopes is represented between $T=$ 0 to 2 MeV. A detailed description of E1 excitations in terms of non-energy-weighted and energy-weighted summations in the low- and high-energy regions is also included. Finally, a summary of the present work is presented in Sec. \ref{sec:summary}. 

\section{\label{sec:Formalism}The finite temperature relativistic quasiparticle random phase approximation}
The fully self-consistent FT-RQRPA based on relativistic energy density functional (REDF) is established in the present work to study the E1 transitions in even-even nuclei. The properties of closed- and open-shell nuclei are described within the finite temperature Hartree-Bardeen–Cooper–Schrieffer (FT-HBCS) framework \cite{Goodman30,PhysRevC.102.065804}, using relativistic density-dependent point coupling DD-PCX interaction \cite{Yuksel99}. This interaction is particularly appropriate for this study because it was adjusted not only to the ground state properties but also to the excitation properties, including dipole polarizability in $^{208}$Pb. In this way, the isovector channel of the interaction is fine-tuned to provide the symmetry energy that is consistent with the experimental data on dipole polarizability.
The point-coupling REDF determined from the Lagrangian density, 
\begin{equation}
\begin{split}
\mathcal{L}_{PC} &= \bar{\psi}(i\gamma \cdot \partial - m)\psi  \\
 &-\frac{1}{2}\alpha_S(\rho) (\bar{\psi}\psi)(\bar{\psi}\psi) 
 -\frac{1}{2}\alpha_V(\rho) (\bar{\psi}\gamma^\mu \psi)(\bar{\psi}\gamma_\mu \psi) \\
 &-\frac{1}{2}\alpha_{TV}(\rho) (\bar{\psi}\vec{\tau}\gamma^\mu \psi)(\bar{\psi}\vec{\tau}\gamma_\mu \psi)
 \\  
&-\frac{1}{2}\delta_S (\partial_\nu \bar{\psi} \psi)(\partial^\nu \bar{\psi} \psi) 
-e\bar{\psi}\gamma \cdot A \frac{1-\tau_3}{2}\psi.
\label{Lagrangian_PC}
\end{split}
\end{equation}
It includes fermion contact interaction terms as isoscalar-scalar, isoscalar-vector, isovector-vector channels.
In addition to the free-nucleon Lagrangian and the point-coupling interaction terms, the model includes the coupling of protons to the electromagnetic field. The derivative term in Eq. (\ref{Lagrangian_PC}) accounts for leading effects of finite-range interactions that are necessary for a quantitative description of nuclear density distribution and radii. For the detailed description of the relativistic point coupling model see Refs. \cite{Niksic2008,Niksic185}.
Directed by the microscopic density-dependence of the vector and scalar self-energies, we use the following practical ansatz for the functional form of the couplings \cite{Niksic2008},
\begin {equation}
\begin{aligned}
 \alpha_{i}(\rho) = a_{i} + (b_{i} + c_{i} x) e^{-d_{i}x}, (i=S, V, TV)
\end{aligned}
\end {equation}
where $x = \rho/\rho_{sat.}$ and $\rho_{sat.}$ represents the nucleon density at saturation point in symmetric nuclear matter. The parameters for each channel $i=S,V,TV$ are represented as $a_{i}$, $b_{i}$, $c_{i}$, $s_{i}$, and $\delta_{S}$ denotes the strength of isoscalar-scalar derivative term. Based on the Dirac-Brueckner calculations of asymmetric nuclear matter \cite{Jong1998}, the number of free parameters is reduced to ten \cite{Niksic2008}. For open-shell nuclei, the theory framework is extended to include a separable form of pairing interaction introduced in Ref. \cite{Tian80}. At finite temperature, the occupation probabilities of single particle states are given by
\begin{equation}
n_i=v_{i}^{2}(1-f_{i})+u_{i}^{2}f_{i},
\end{equation} 
where $u_i$ and $v_i$ are the BCS amplitudes. The temperature dependent Fermi-Dirac distribution function is defined as
\begin{equation}
\label{fermi_dirac}
f_{i}=[1+exp(E_{i}/k_{B}T)]^{-1},
\end{equation}
where $T$  and $k_{B}$ are temperature and Boltzmann constant, respectively. $E_{i}$ is the quasiparticle (q.p.) energy of a state and is calculated using $E_i=\sqrt{(\varepsilon_i-\lambda_q)^2+\Delta_i^2}$ relation, where $\varepsilon_i$ represents the single-particle energies and $\lambda_q$ denotes chemical potentials for either proton or neutron states. $\Delta_i$ indicates the pairing gap of the given state. A sharp pairing phase transition is anticipated from super-fluid state to the normal state at critical temperatures ($T_c$) \cite{Goodman30,PhysRevC.88.034308, PhysRevC.93.024321}. 
The values of $T_c$ for considered open-shell Ca and Sn nuclei are calculated using FT-HBCS and presented in Table \ref{table1:Tc}.

\begin{table}[ht] \renewcommand{\arraystretch}{1.1}
\tabcolsep 0.35cm
\caption{The critical temperature ($T_c$) values for pairing phase transition in open-shell Ca and Sn nuclei.} 
\centering 
\begin{tabular}{c c c c } 
\hline \\[-1.0em]
 Nucleus & $T_c$ [MeV] &  Nucleus & $T_c$ [MeV]  \\ [1ex] 
\hline \\[-1.0em] 
$^{44}$Ca & 0.862 & $^{108}$Sn & 0.872 \\ 
$^{52}$Ca & 0.528 & $^{116}$Sn & 0.834\\ 
$^{56}$Ca & 0.743 & $^{124}$Sn & 0.764 \\ 
 &  & $^{140}$Sn & 0.644 \\ 
\hline \\ [-1.ex]
\end{tabular}
\label{table1:Tc} 
\end{table}

To analyze the isovector dipole excitations in nuclei, the FT-RQRPA framework is applied on top of the FT-HBCS. The description of excitation operator and the derivation of the expressions for the matrix elements of the finite temperature non-relativistic quasiparticle random phase approximation, based on Skyrme functionals, are given in Refs. \cite{Yuksel96,Yuksel55}. Here, we consider the formalism for the FT-RQRPA based on relativistic point-coupling interaction discussed above. 
The finite temperature non-charge exchange RQRPA matrix is given by
\begin{equation}
\left( { \begin{array}{cccc}\label{eq:qrpa}
 \widetilde{C} & \widetilde{a} & \widetilde{b} & \widetilde{D} \\
 \widetilde{a}^{+} & \widetilde{A} & \widetilde{B} & \widetilde{b}^{T} \\
-\widetilde{b}^{+} & -\widetilde{B}^{\ast} & -\widetilde{A}^{\ast}& -\widetilde{a}^{T}\\
-\widetilde{D}^{\ast} & -\widetilde{b}^{\ast} & -\widetilde{a}^{\ast} & -\widetilde{C}^{\ast}
 \end{array} } \right)
 \left( {\begin{array}{cc}
\widetilde{P}  \\
\widetilde{X }  \\
\widetilde{Y}  \\
\widetilde{Q} 
 \end{array} } \right)
 = E_{w}
  \left( {\begin{array}{cc}
\widetilde{P}  \\
\widetilde{X}  \\
\widetilde{Y}  \\
\widetilde{Q} 
\end{array} } \right), \end{equation}
where $ E_{w}$ denotes the excitation energies and eigenvectors $\widetilde{P}, \widetilde{X}, \widetilde{Y}, \widetilde{Q}$ are read as
\begin{gather}
\begin{aligned}
\widetilde{X}_{ab}=X_{ab}\sqrt{1-f_{a}-f_{b}},
\end{aligned} \\
\begin{aligned}
\widetilde{Y}_{ab}=Y_{ab}\sqrt{1-f_{a}-f_{b}},
\end{aligned} \\
\begin{aligned}
\widetilde{P}_{ab}=P_{ab}\sqrt{f_{b}-f_{a}},
\end{aligned}\\
\begin{aligned}
\widetilde{Q}_{ab}=Q_{ab}\sqrt{f_{b}-f_{a}}.
\end{aligned}
\end{gather}
The FT-RQRPA matrices are diagonalized in a self-consistent way, providing a state-by-state analysis for each excitation. The $T$-dependent matrix elements are given as,
\begin{align}
\begin{split}
\widetilde{A}_{abcd}=&\sqrt{1-f_{a}-f_{b}} A'_{abcd}\sqrt{1-f_{c}-f_{d}}\\
&+(E_{a}+E_{b})\delta_{ac}\delta_{bd} \label{eq:temp},
\end{split}\\
\begin{split}
\widetilde{B}_{abcd}=&\sqrt{1-f_{a}-f_{b}} B_{abcd}\sqrt{1-f_{c}-f_{d}}, 
\end{split}\\
\begin{split}
\widetilde{C}_{abcd}=&\sqrt{f_{b}-f_{a}} C'_{abcd}\sqrt{f_{d}-f_{c}}\\
&+(E_{a}-E_{b})\delta_{ac}\delta_{bd},
\end{split}\\
\begin{split}
\widetilde{D}_{abcd}=&\sqrt{f_{b}-f_{a}} D_{abcd}\sqrt{f_{d}-f_{c}}, 
\end{split}\\
\begin{split}
\widetilde{a}_{abcd}=&\sqrt{f_{b}-f_{a}} a_{abcd}\sqrt{1-f_{c}-f_{d}}, 
\end{split}\\
\begin{split}
\widetilde{b}_{abcd}=&\sqrt{f_{b}-f_{a}} b_{abcd}\sqrt{1-f_{c}-f_{d}}, 
\end{split}\\
\begin{split}
\widetilde{a}_{abcd}^{+}=&\widetilde{a}_{abcd}^{T}=\sqrt{f_{d}-f_{c}} a_{abcd}^{+}\sqrt{1-f_{a}-f_{b}}, 
\end{split}\\
\begin{split}
\widetilde{b}_{abcd}^{T}=&\widetilde{b}_{abcd}^{+}=\sqrt{f_{d}-f_{c}} b_{abcd}^{T}\sqrt{1-f_{a}-f_{b}},\label{eq:temp1}
\end{split}
\end{align}
where $E_{a(b)}$ is the quasiparticle energy of the states obtained from the FT-HBCS results. The $A$ and $B$ matrix elements contribute both at zero and finite temperature as these describe the effects of the excitations of quasiparticle pairs. The other elements of the FT-RQRPA matrix start to contribute as temperature increases, as they are influenced by the changing occupation factors. Note that the diagonal part of the matrix includes both ($E_{a}+E_{b}$) and ($E_{a}-E_{b}$) configuration energies, and the latter starts to contribute with increasing temperature and mainly impacts the low-energy part of the excitation spectrum. The full expressions of all the matrices is given in Refs. \cite{Yuksel96,Sommermann}. 
\begin{figure*}[!ht]
\includegraphics[width=\linewidth,clip=true]{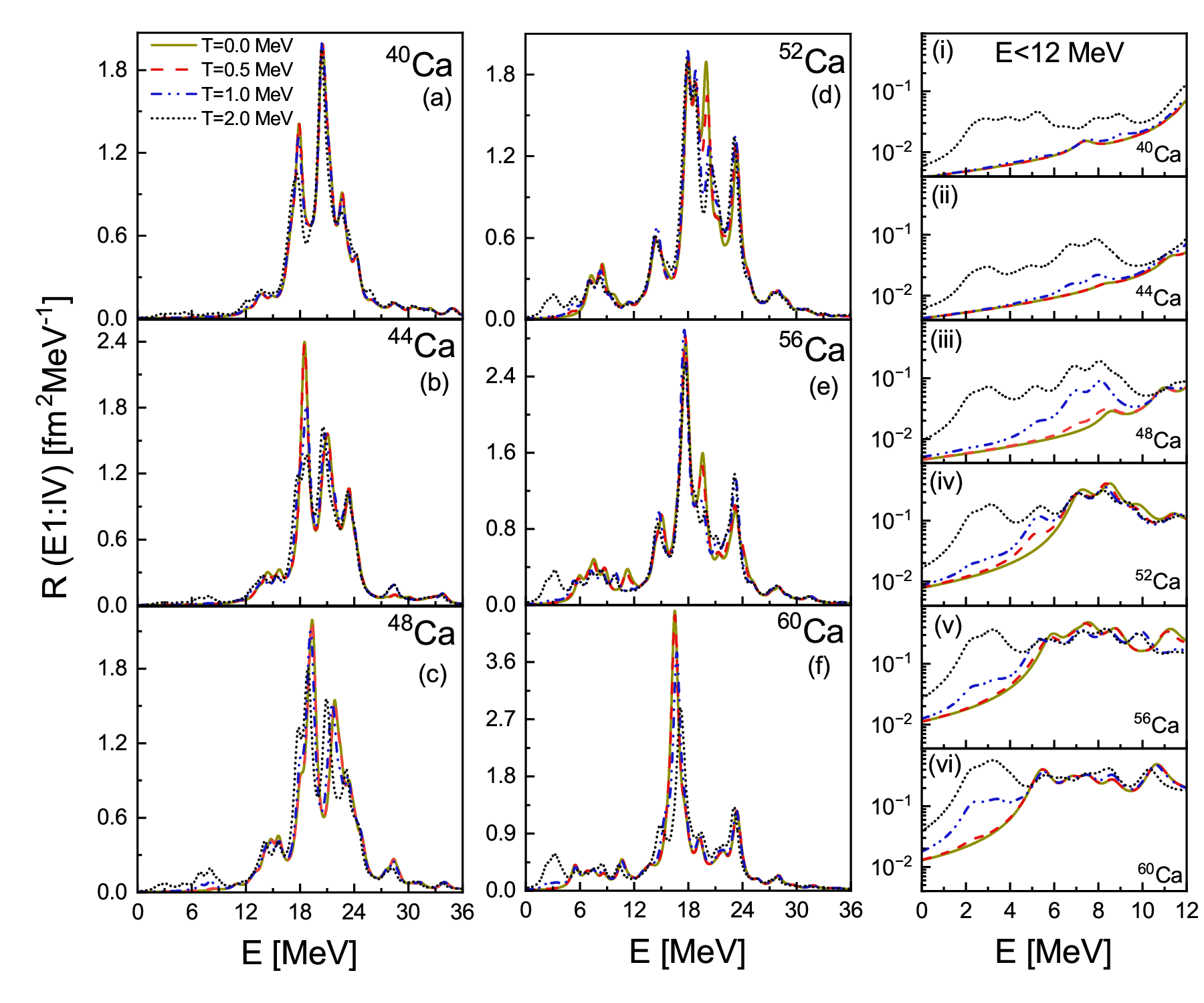}
 \caption{The isovector dipole strength distributions for $^{40-60}$Ca isotopic chain [Panels [(a)-(f)]. The calculations are performed using the FT-RQRPA with DD-PCX interaction at temperatures $T=$ 0, 0.5, 1, and 2 MeV. The low-energy part of the E1 strength for $E<$ 12 MeV is also displayed in logarithmic scale in panels (i)-(vi) (color online).}
  \label{fig1}
\end{figure*}

Our calculations are fully self-consistent: the same REDF and separable pairing approach is employed both in the FT-HBCS calculations and in the FT-RQRPA residual interaction. 
At finite temperature, the reduced transition probability is calculated as
\begin{widetext}
\begin{equation}
\begin{split}
B(EJ,\widetilde0\rightarrow w)=\bigl|\langle w ||\hat{F}_{J}||\widetilde0\rangle \bigr|^{2}
&=\biggl|\sum_{c\geq d}\Big\{(\widetilde{X}_{cd}^{w} + (-1)^{j_{c}-j_{d}+J}\widetilde{Y}_{cd}^{w})(u_{c}v_{d}+(-1)^{J}v_{c}u_{d})\sqrt{1-f_{c}-f_{d}} \\
&+(\widetilde{P}_{cd}^{w}+(-1)^{j_{c}-j_{d}+J}\widetilde{Q}_{cd}^{w})(u_{c}u_{d}-(-1)^{J}v_{c}v_{d})\sqrt{f_{d}-f_{c}}\Big\}\langle c ||\hat{F}_{J}||d\rangle\biggr|^{2}.
\end{split}
\label{bel}
\end{equation}
\end{widetext}
$\hat{F}_{J}$ is the transition operator of the relevant excitation. In this work, the isovector E1 operator is used to calculate electric transition strength distributions \cite{paar67}. In Eq. (\ref{bel}), $|w\rangle$ is the excited state and $|\widetilde0\rangle$ is the correlated FT-RQRPA vacuum state. It is also interesting to consider contribution of a particular proton or neutron configuration in the total E1 transition strength at given excitation energy $E_w$,

\begin{equation}\label{Part.Contri}
B (EJ, E_{w})=\big|\sum_{cd}\left(b^{\pi}_{cd}(E_w)+b^{\nu}_{cd}(E_w)\right)\big|^2.
\end{equation}

Here, $b^{\pi}_{cd}$($E_w$) and $b^{\nu}_{cd}(E_w)$ represent the proton ($\pi$) and neutron ($\nu$) partial contributions for a specific configuration $cd$. In the present work, the quasiparticle cut-off energies for the configuration space in the FT-RQRPA are selected to provide a sufficient convergence in the E1 excitation strength ($E_{cut}$ = 100 MeV). We performed the calculations with the assumption of spherical symmetry, and 20 oscillator shells are used in the FT-HBCS calculations. Then, the discrete FT-RQRPA spectrum is averaged with a
Lorentzian of $\Gamma=1.0$ MeV width using
\begin{equation}\label{lorentz}
R(EJ,E_{w})=\sum_{w}\frac{1}{2\pi}\frac{\Gamma}{(E-E_{w})^{2}-\Gamma^{2}/4}B(EJ,\widetilde0\rightarrow w).
\end{equation}

\begin{table}[ht!] \renewcommand{\arraystretch}{1.2}
\tabcolsep 0.25cm
\caption{The partial contributions $b_{2qp}^{\pi(\nu)}$ [fm] of major proton ($\pi$) and neutron ($\nu$) transitions to the main high-energy E1 peak of the $^{56}$Ca nucleus are evaluated at $T=$ 0 and 2 MeV. The total E1 strength is obtained by summing over all proton and neutron configurations (see Eq. (\ref{Part.Contri})), also including those not listed in the table.}
\centering 
\begin{tabular}{l c c} 
\hline \\[-1.0em]
   Configuration                         & $T= 0$ MeV    & $T = 2$ MeV \\
                                         & $E=$ 17.76 MeV & $E=$ 17.59 MeV     \\ \hline
                                          
$\nu$(1f$_{7/2}$$\rightarrow$1g$_{9/2}$) & -0.224        &0.219         \\
$\nu$(1f$_{7/2}$$\rightarrow$3d$_{5/2}$) & -0.036        &-0.095        \\
$\nu$(1f$_{7/2}$$\rightarrow$2g$_{9/2}$) & 0.052         &-0.058        \\
$\nu$(1d$_{3/2}$$\rightarrow$1f$_{5/2}$) & -0.188        &0.293         \\
$\nu$(1f$_{7/2}$$\rightarrow$1g$_{7/2}$) & -0.092        &0.069         \\ 
$\nu$(2s$_{1/2}$$\rightarrow$2p$_{1/2}$) & -0.055        &0.027              \\
$\nu$(1f$_{5/2}$$\rightarrow$2g$_{7/2}$) & -0.027        &0.043         \\ \hline
$\pi$(1d$_{3/2}$$\rightarrow$1f$_{5/2}$) & -0.592        & 0.396        \\
$\pi$(1d$_{5/2}$$\rightarrow$1f$_{7/2}$) & -0.402        & 0.319        \\
$\pi$(1d$_{5/2}$$\rightarrow$2p$_{3/2}$) & 0.199         & -0.054         \\
$\pi$(1d$_{3/2}$$\rightarrow$2p$_{1/2}$) & -0.080        & 0.031         \\
$\pi$(2s$_{1/2}$$\rightarrow$2p$_{3/2}$) & -0.083        & 0.055         \\
$\pi$(1p$_{1/2}$$\rightarrow$1d$_{3/2}$) &               & 0.117        \\
$\pi$(1f$_{7/2}$$\rightarrow$2d$_{5/2}$) &               &-0.119        \\\hline
total $B(E1)$ [fm$^2$]               & 2.463         &1.760         \\
\hline \\ [-1.ex]
\end{tabular}
\label{table2} 
\end{table}
\begin{table}[ht!] \renewcommand{\arraystretch}{1.2}
\tabcolsep 0.25cm
\caption{The same as in Table \ref{table2} but for the selected low-energy E1 excitations for $E <$ 5 MeV at $T=2$ MeV.}
\centering 
\begin{tabular}{c l c c} 
\hline \\[-1.0em]
Energy     &    Configuration      & $b_{2qp}^{\pi(\nu)}$     & total $B(E1)$   \\            
$[$MeV$]$ &                              & [fm]                      & [fm$^2$]       \\ \hline
1.88 & $\nu$(2g$_{9/2}$$\rightarrow$1h$_{11/2}$)  & -0.142       & 0.022        \\
2.19 & $\nu$(3s$_{1/2}$$\rightarrow$3p$_{3/2}$)   & 0.331        & 0.122        \\
2.27 & $\nu$(2f$_{7/2}$$\rightarrow$2g$_{9/2}$)   &  0.255       & 0.064        \\
2.42 & $\nu$(3s$_{1/2}$$\rightarrow$3p$_{1/2}$)   & -0.233       & 0.054       \\
2.73 & $\nu$(1h$_{11/2}$$\rightarrow$1i$_{13/2}$) & -0.116       & 0.014        \\
2.96 & $\nu$(2d$_{3/2}$$\rightarrow$2f$_{5/2}$)   & 0.344        & 0.124        \\
3.26 & $\nu$(2d$_{5/2}$$\rightarrow$2f$_{7/2}$)   &  0.453       & 0.212        \\
3.42 & $\nu$(3p$_{3/2}$$\rightarrow$4s$_{1/2}$)   &  0.134       & 0.018        \\
3.46 & $\nu$(1g$_{7/2}$$\rightarrow$1h$_{9/2}$)   & -0.207      & 0.042        \\ 
3.59 & $\nu$(2g$_{9/2}$$\rightarrow$2h$_{11/2}$)  & 0.186        & 0.045       \\
4.18 & $\nu$(3p$_{1/2}$$\rightarrow$3d$_{3/2}$)   & 0.146        & 0.020        \\
4.42 & $\nu$(2f$_{5/2}$$\rightarrow$2g$_{7/2}$)   & 0.129        & 0.015        \\
\hline \\ [-1.ex]
\end{tabular}
\label{table3}
\end{table}
\begin{figure*}[!ht]
\includegraphics[width=\linewidth,clip=true]{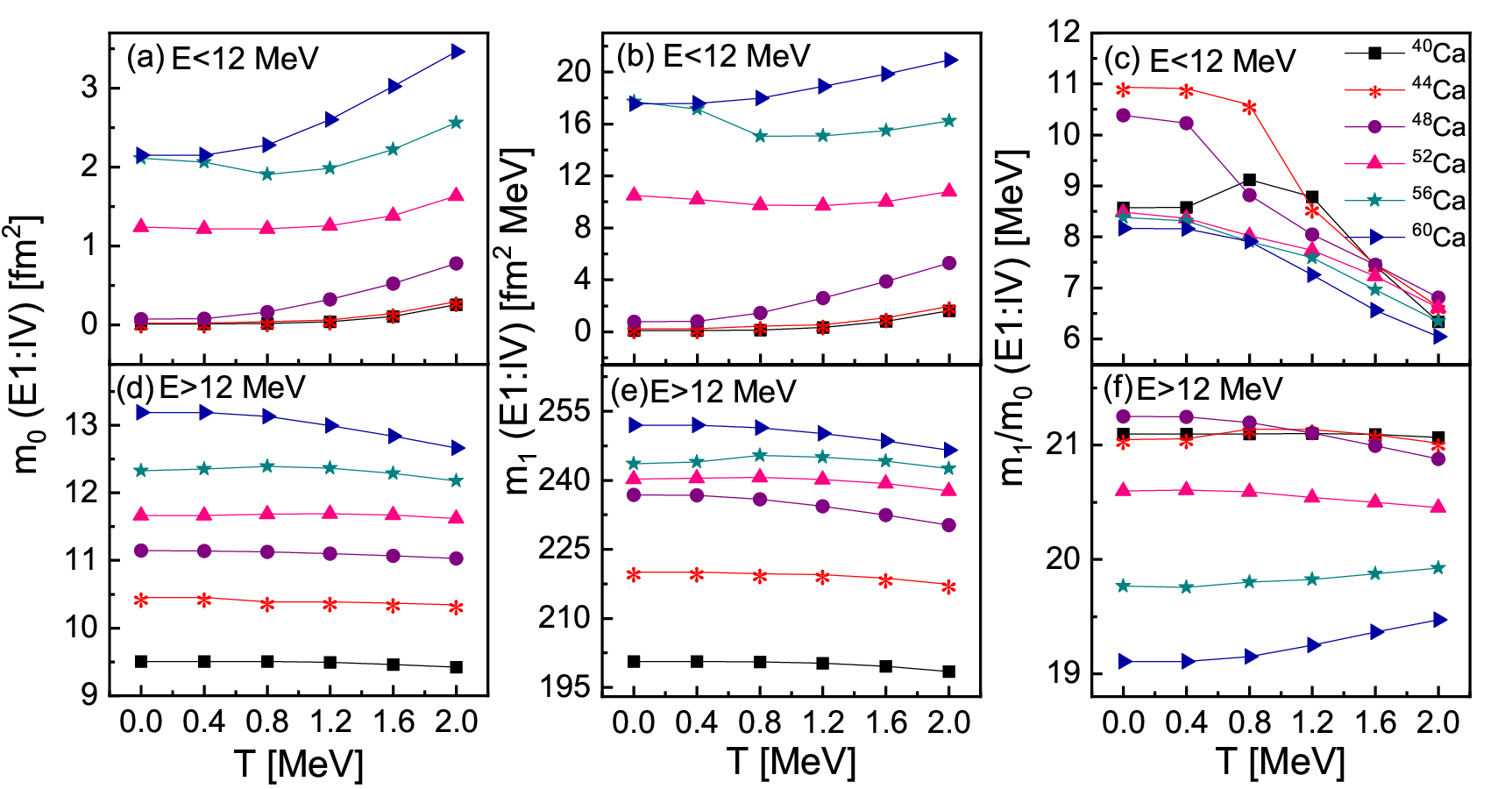}
  \caption{The $m_0$ and $m_1$ moments of the strength function and the centroid energy $m_1/m_0$ of the isovector dipole response of the selected Ca nuclei in the low-energy region for $E < 12$ MeV (panels (a)-(c)) and the high-energy region for $E > 12$ MeV (panels (d)-(f)) as a function of temperature.}
  \label{fig2}
\end{figure*}
\section{\label{sec:results} Results}
In this section, the FT-RQRPA introduced in Sec. \ref{sec:Formalism} is employed in studies of electric dipole (E1) response at finite temperature of closed- and open-shell $^{40-60}$Ca and $^{100-140}$Sn isotopes. In particular, the evolution of the E1 response with an increase in temperature is considered. The characteristic behavior of low- and high-energy response is investigated with the variation of temperature and neutron number (N) of nucleus. The calculations are performed between $T=$ 0 to 2 MeV, that includes the range both below and above the critical temperature $T_c$. At temperatures higher than critical values (see Table \ref{table1:Tc}), pairing correlations vanish for all open-shell nuclei and do not contribute to the FT-RQRPA matrices.

\subsection{\label{sec:IIIA}Ca isotopes}
In this section, we study the isotopic and temperature dependence of low-energy and high-energy electric dipole transitions for calcium nuclei. In Fig. \ref{fig1}, the isovector E1 transition strength distributions of $^{40-60}$Ca isotopic chain are displayed at temperatures $T=$ 0, 0.5, 1 and 2 MeV. On the right side of Fig. \ref{fig1} [panels (i)-(vi)], the low-energy strength is also displayed on a logarithmic scale to provide a better insight into the changes occurring in the low-energy region. At $T =$ 0 MeV, it is observed that low-energy excited states begin to emerge, and the strength increases as the neutron number of Ca isotopes increases. Low-energy excited states have been obtained for the neutron-rich $^{52,56,60}$Ca nuclei with energies $E$ $<$ 12 MeV, which are also referred to as the pygmy dipole strength \cite{Vretenar692, SAVRAN2013210, BRACCO2019360, Paar_2007}. At low temperatures and $T=$ 0.5 MeV, the dipole strength almost does not change. By increasing the temperature to 2 MeV, the high-energy E1 strength redistributes over the main peaks, and the excited states start shifting slightly towards lower energies, as depicted in Fig. \ref{fig1}. The impact of temperature becomes more prominent in the low-energy region. At $T =$ 1 MeV, new low-energy states start to emerge for $E<$ 12  MeV. When the temperature is raised to $T =$ 2 MeV, its influence becomes even more pronounced in the low-energy region of neutron-rich nuclei, and leads to the emergence of new low-energy excited states with significant strength. At finite temperatures, the promotion of nucleons to higher states results in an increase (or decrease) in the occupation probabilities of states above (or below) the Fermi level, thus leading to the broadening of the Fermi surface. Consequently, thermal unblocking effects play a crucial role in giving rise to new excitation channels, particularly in the low-energy region of the electric dipole response.

For more detailed analysis, the proton ($\pi$) and neutron ($\nu$) transitions with the largest partial contributions $b^{\pi,\nu}_{cd}$ to the FT-RQRPA transition strength (see Eq. (\ref{Part.Contri})) of the main E1 high-energy peak ($E\approx$ 17.5 MeV) in $^{56}$Ca are listed in Table \ref{table2} at $T=$ 0 and 2 MeV. It is found that the high-energy peak is composed of the coherent contributions from both neutron and proton transitions in both cases. By inspecting E1 partial contributions, we observe that the number of neutron and proton configurations and their amplitudes attributing to the major high-energy excitation altered slightly at $T=$2 MeV. This signifies that the GDR region of E1 strength experiences subtle modifications at high temperatures.

Table \ref{table3} clearly shows the emergence of new low-energy transitions in $^{56}$Ca for $E<$ 5 MeV at $T=$ 2 MeV. Notably, the structure of these low-energy peaks is primarily determined by the single neutron configuration, which mainly comes from the high-energy continuum states. Given the number of transitions that contribute to these states and their strength, the newly formed low-energy states do not exhibit the same collective behavior observed in the case of the high-energy region.
In neutron-rich nuclei, neutrons can easily populate the continuum states at high temperatures. Therefore, the contributions coming from the continuum states play a major role in the formation of low-energy states at finite temperatures. Increasing the temperature further and above $T>$ 2 MeV can lead to the formation of neutron vapor and unphysical strength in the low-energy region. Hence, special care must be taken either by limiting the temperature or by properly taking the continuum into account in the calculations. In our calculations, we constrained the temperature to be up to $T=2$ MeV to mitigate this issue.

\begin{figure*}[ht!]
\includegraphics[width=\linewidth,clip=true]{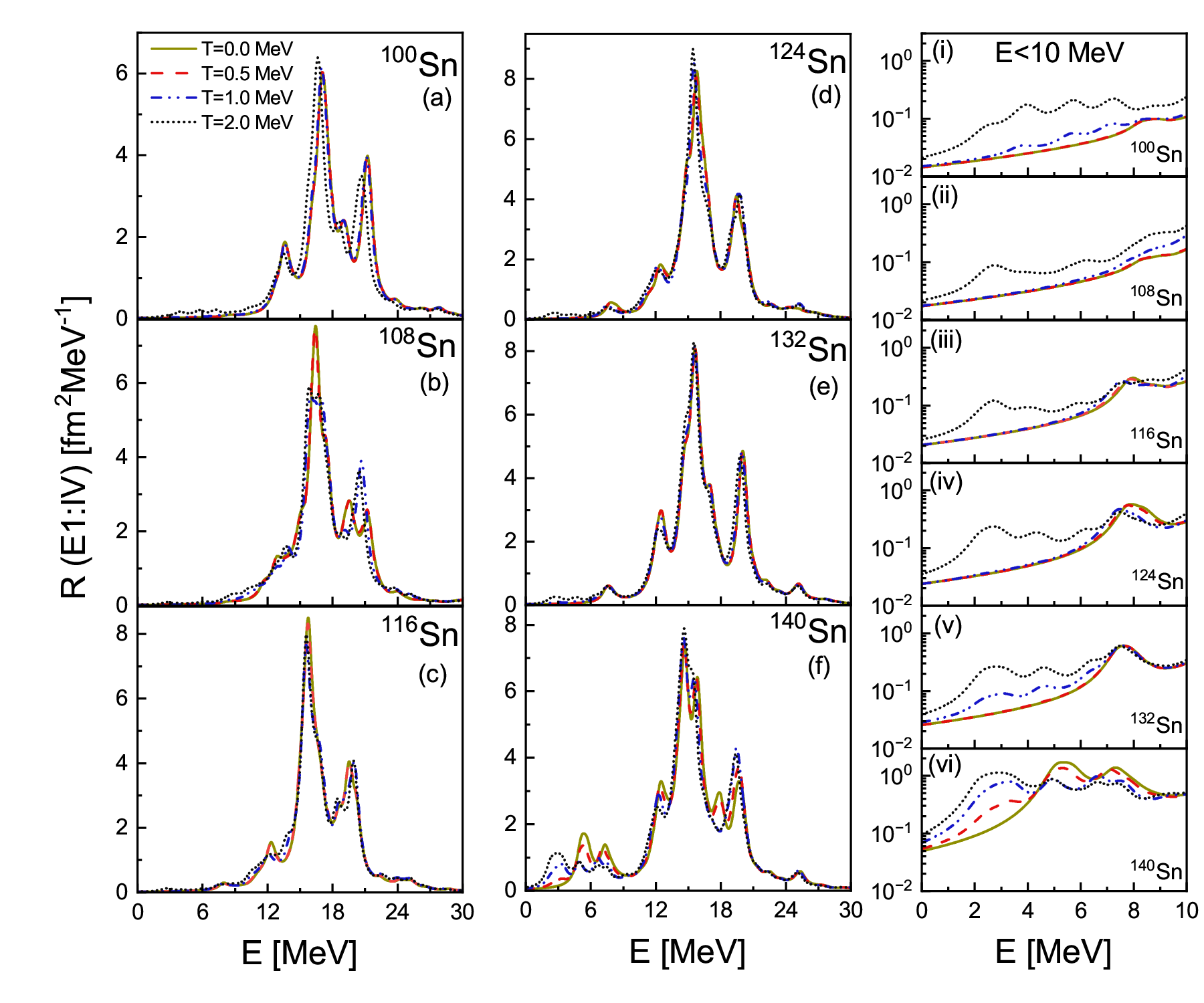}
  \caption{The same as in Fig. \ref{fig1} but for the selected nuclei between $^{100-140}$Sn.}
  \label{fig3}
\end{figure*}

Figure \ref{fig2} shows the variation in total dipole transition strength moments both below [upper panels (a-c)] and above [lower panels (d-f)] 12 MeV for $^{40-60}$Ca nuclei. The total transition strength $m_0$, energy weighted strength $m_1$ and centroid energy $E_c=m_1/m_0$ are displayed with increasing temperature. As shown in Fig. \ref{fig1}, the energy $E =$ 12 MeV can be used to distinguish between the low-energy pygmy dipole strength from the high-energy Giant Dipole Resonance region in all Ca isotopes. Hence, this energy value serves as a suitable choice to delineate the two regions for our analysis.

Let us first consider the low-energy region for $E<$ 12 MeV at zero temperature. It can be seen that the $m_0$ (a) and $m_1$ moments (b) display an increase, while the centroid energy (c) decreases with increasing neutron number, as expected. By increasing the temperature, the moments of the strength function also undergo changes. Up to $T =$ 0.4 MeV, they do not exhibit sensitivity to the temperature and remain almost constant for all nuclei. As the temperature increases further, specifically beyond $T>$ 0.8 MeV, the $m_0$ and $m_1$ moments display an increase. For $^{56}$Ca, we observe a slight decrease in the $m_0$ and $m_1$ moments between $0.4 < T < 0.8$ MeV, which is attributed to rapid changes in the pairing properties around the critical temperatures. It is also seen that the centroid energy $E_c$ (c) of Ca isotopes in the low-energy region remains almost constant up to $T=$0.4 MeV and then starts to decrease rapidly with increasing temperature. Only for $^{40}$Ca, we observe a slight increase in the centroid energy around $T=$ 0.8 MeV, which is due to the energy range  considered in the study. At higher temperatures, the centroid energy also decreases for this nucleus. It is observed that neutron-rich isotopes $^{48-60}$Ca are more sensitive to changes in temperature, and this increase occurs at a more rapid rate for the $m_0$ moment due to the increasing contribution of continuum states to the low-energy region at higher temperatures, as mentioned above. 

In the high-energy or giant dipole resonance region ($E >$ 12 MeV) and at zero temperature, the $m_0$ and $m_1$ moments increase, while the centroid energy decreases with increasing neutron number, as expected. With increasing temperature, the $m_0$ and $m_1$ moments exhibit a relatively gradual change. For all Ca nuclei, the $m_0$ and $m_1$ moments remain almost constant up to $T=0.8$ MeV, beyond which they exhibit a very slight decrease because the GDR strength distributions are slightly modified at the considered temperature range. The centroid energy also displays the same behaviour and  remains almost stable below $T=0.8$ MeV for all Ca nuclei. At higher temperatures, the centroid energy of $^{40}$Ca remains almost stable, while it starts to decrease (increase) slightly for the neutron-rich $^{48}$Ca ($^{60}$Ca). Since these nuclei do not have pairing correlations, the changes in the dipole response are mainly related to two factors. First, there is the softening of the repulsive residual particle-hole (ph) interaction. Second, there are changes in the configuration energies, along with the contribution of new configurations due to the thermal unblocking effect at finite temperatures. For open shell nuclei, $^{44}$Ca exhibit a modest decrease, and a small increase is obtained for $^{56}$Ca above the critical temperatures. We also note that the centroid energy decreases slightly for $^{52}$Ca.

\begin{figure*}[!ht]
\includegraphics[width=1.0\linewidth,clip=true]{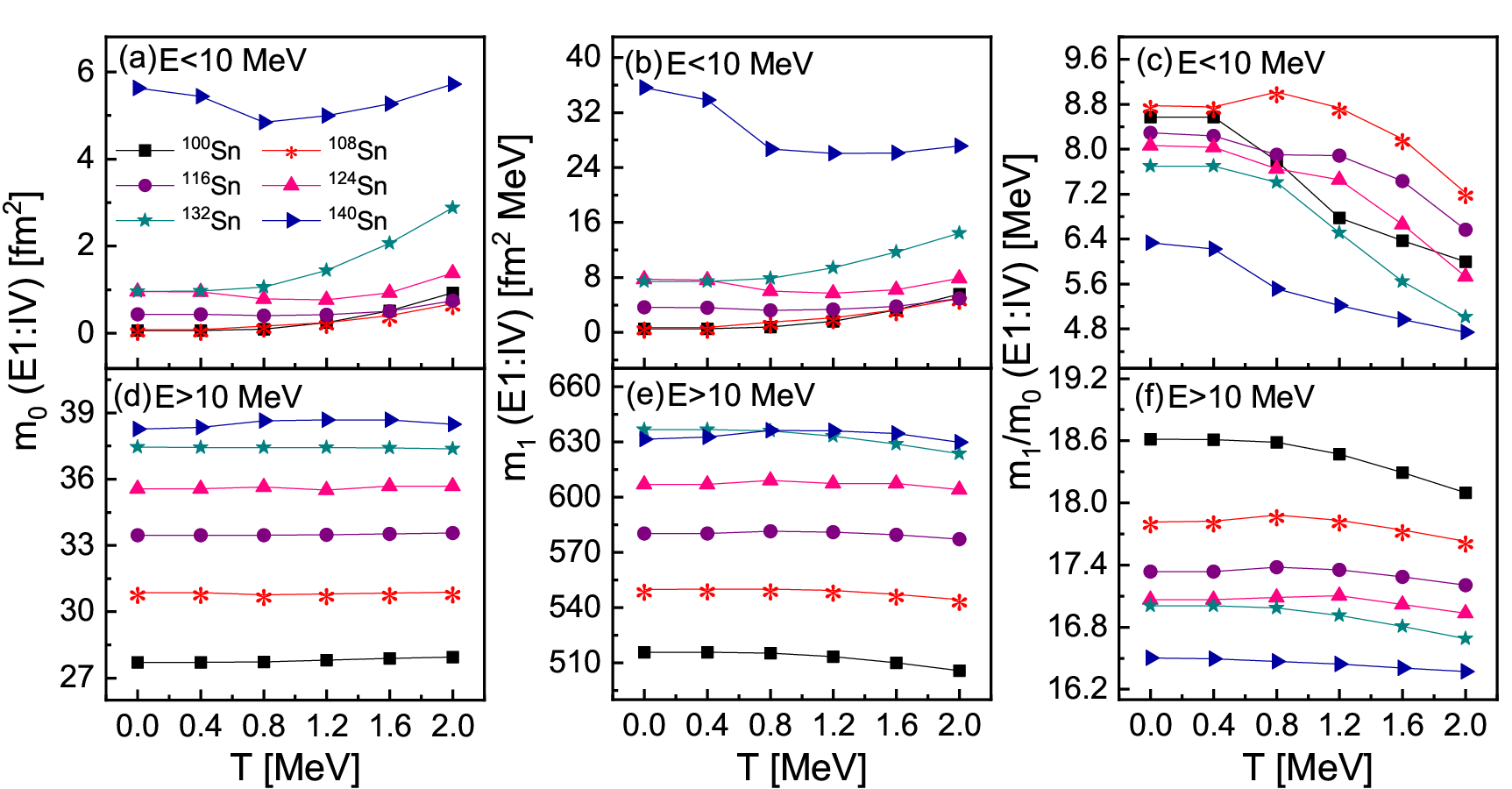}
 \caption{The $m_0$ and $m_1$ moments of the strength function and the centroid energy $m_1/m_0$ of the isovector dipole response of the selected Sn nuclei in the low-energy region for $E < 10$ MeV (panels (a)-(c)) and the high-energy region for $E > 10$ MeV (panels (d)-(f)) as a function of temperature.}
  \label{fig4}
\end{figure*}

\subsection{\label{sec:IIIB}Sn isotopes}
In the next step, we employ the FT-RQRPA to calculate the isovector E1 strength for heavier mass nuclei, specifically $^{100-140}$Sn, at temperatures of $T =$ 0, 0.5, 1, and 2 MeV. The results are depicted in Fig. \ref{fig3}. At $T =$ 0 and 0.5 MeV, the isotopic dependence of the low-energy energy modes for $E$$<$10 MeV in Sn isotopes is clearly visible from the figure. The E1 response function of $^{100, 108}$Sn displays negligible contribution in the low-energy region. However, the low-energy strength starts to become visible when one moves from $^{116}$Sn to $^{140}$Sn. In $^{140}$Sn nucleus, a significant contribution of low-energy E1 excitations is observed between 4-9 MeV due to the larger neutron content. 

At finite temperatures, the highly collective Giant Dipole Resonance (GDR) region exhibits only minor changes in terms of the strength and excitation energies. By increasing temperature further, at $T =$ 1 and 2 MeV, newly emerged low-energy peaks can be observed at $E$ $<$ 5 MeV for neutron-rich $^{124-140}$Sn isotopes due to thermal unblocking. This is depicted separately on the right side in [panels (i)-(vi)] of Fig. \ref{fig3}. The impact of temperature is most prominent in the low-energy region of the $^{140}$Sn nucleus due to its excessive neutron content. Table \ref{table4} represents the partial contributions of particle configurations to the newly emerged low-energy E1 peaks at $E<$ 5 MeV in $^{140}$Sn nucleus at $T =$ 2 MeV. Similar to the findings in Sec \ref{sec:IIIA} for $^{56}$Ca, these low-energy peaks are predominantly attributed to single neutron transitions, primarily caused by thermal unblocking effects.

\begin{table}[ht] \renewcommand{\arraystretch}{1.2}
\tabcolsep 0.25cm
\caption{The same as described in the caption for Table \ref{table2}, but for the low-energy E1 excitations below $E$$=$5 MeV in $^{140}$Sn nucleus at $T=$ 2 MeV.}
\begin{tabular}{c l c c} 
\hline \\[-1.0em]
Energy     &    Configuration      & $b_{2qp}^{\pi(\nu)}$     & total $B(E1)$  \\    
$[$MeV$]$     &                      & [fm]                    &  [fm$^2$]  \\ \hline
1.52 & $\nu$(3p$_{3/2}$$\rightarrow$4s$_{1/2}$) & -0.244         & 0.063        \\
1.91 & $\nu$(2h$_{11/2}$$\rightarrow$2i$_{13/2}$) & -0.253       & 0.069        \\
2.16 & $\nu$(3d$_{5/2}$$\rightarrow$3f$_{7/2}$) &  0.548         & 0.345        \\
2.20 & $\nu$(4s$_{1/2}$$\rightarrow$4p$_{3/2}$) & -0.366         & 0.138        \\
2.36 & $\nu$(3d$_{3/2}$$\rightarrow$3f$_{5/2}$) &  0.443         & 0.202        \\
2.69 & $\nu$(3p$_{3/2}$$\rightarrow$4s$_{1/2}$) & -0.433         & 0.226        \\ 
2.89 & $\nu$(2g$_{9/2}$$\rightarrow$2h$_{11/2}$) & 0.466         & 0.227        \\
2.95 & $\nu$(2f$_{5/2}$$\rightarrow$3d$_{3/2}$) & -0.383         & 0.172        \\
3.16 & $\nu$(3p$_{1/2}$$\rightarrow$3d$_{3/2}$) & 0.486          & 0.237        \\ 
3.45 & $\nu$(3p$_{3/2}$$\rightarrow$3d$_{5/2}$) & -0.689         & 0.485        \\     
4.89 & $\nu$(2f$_{7/2}$$\rightarrow$3d$_{5/2}$) & 0.546          & 0.543        \\    
     & $\nu$(1h$_{11/2}$$\rightarrow$1i$_{13/2}$) & -0.149       &              \\  
     & $\nu$(1g$_{9/2}$$\rightarrow$1h$_{11/2}$) & 0.208         &              \\  \hline \\ [-1.ex]
\end{tabular}
\label{table4}
\end{table}

For further analysis of E1 excitations in Sn nuclei, the moments of the strength distribution $m_0$, $m_1$ and the centroid energy $E_c=m_1/m_0$ are plotted in Fig. \ref{fig4} as a function of temperature $T$ in the low $E$$<$10 MeV [upper panel (a-c)] and high $E$$>$10 MeV [lower panel (d-f)] energy regions. The separation energy of 10 MeV is selected for demonstration purposes because it reasonably well separates the FT-RQRPA low-energy E1 strength from the GDR one for all Sn isotopes (see Fig. \ref{fig3}). In the low-energy region ($E$$<$10 MeV), the $^{100-116}$Sn nuclei exhibit rather small increase in the total B(E1) strength ($m_0$) and energy-weighted ($m_1$) moment as the temperature rises, in contrast to more neutron-rich $^{124-140}$Sn nuclei. 
We also found that the $m_0$ and $m_1$ moments of the $^{140}$Sn nucleus initially decrease and then increase with increasing temperature. This is because of the rapid changes in the pairing properties around critical temperature. Similar to the findings in Sec. \ref{sec:IIIA}, the results show that neutron-rich nuclei are more prone to develop low-energy transitions at higher temperatures. Fig. \ref{fig4}(c) shows that the centroid of PDR energies start to shift downwards rapidly at higher temperatures for Sn nuclei. In Figs. \ref{fig4}(d-f), the isotopic and temperature dependence of giant dipole resonance ($E$$>$10 MeV) of Sn nuclei is explored in terms of $m_0$, $m_1$ and $m_1/m_0$ moments between $T=$ 0 to 2 MeV. It is evident that total E1 strength of Sn nuclei remains almost constant and $m_1$ moment decreases slightly with increasing temperature.
The slight decrease in centroid energy of high-energy resonance ($E$$>$10 MeV) indicates that the E1 GDR strength shifted moderately downward with increasing temperature. However, compared to the low-energy strength, the GDR centroid shift is considerably smaller. In conclusion, temperature effects play a crucial role in the emergence of new low-energy E1 excitations, particularly in neutron-rich nuclei. Conversely, temperature has only a minor impact on the E1 response in the high-energy region. The high collectivity in the GDR region is moderately influenced by thermal modifications of single-particle states around the Fermi level. 
\begin{figure}[!ht]
\includegraphics[width=\linewidth,clip=true]{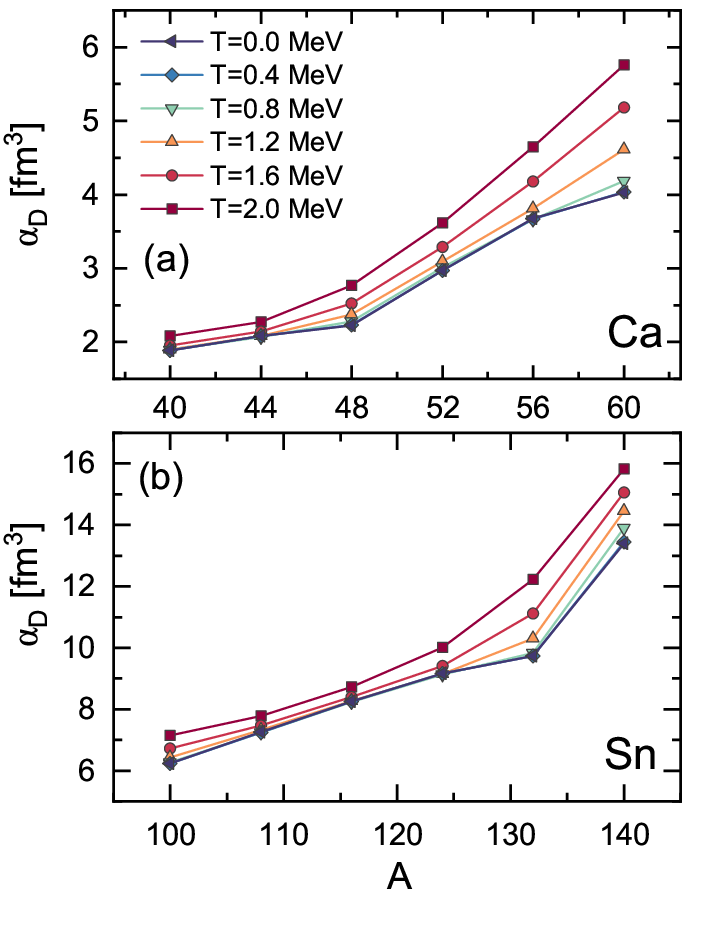}
  \caption{Dipole polarizability $\alpha_{D}$ of $^{40-60}$Ca [panel (a)] and $^{100-140}$Sn [panel (b)] isotopes as a function of atomic mass number (A). The calculations are performed at finite temperatures ranging from $T=0$ to 2 MeV.}
  \label{fig5}
\end{figure}

\subsection{Dipole polarizability at finite temperature}\label{sec:IIIC}
The electric dipole polarizability ($\alpha_D$) is another measurable quantity related to the response of the nucleus when it is subjected to an external electric field \cite{Tamii2011}. Due to the isospin dependence of $\alpha_D$, it provides new insights into the properties of neutron-rich matter and constrains the equation of state (EOS) \cite{Reinhard2010}. $\alpha_D$ is highly sensitive to the density dependence of the symmetry energy and is directly proportional to the inverse energy-weighted E1 sum rule $m_{-1}$. It is calculated by \cite{Tamii2011,Bohigas102,Vretenar85},

\begin{equation}
    \alpha_D= \frac{8\pi e^2}{9}m_{-1}(E1).
\end{equation}

In the previous discussion, we show that E1 moments exhibit a temperature dependence in both low- and high-energy regions, although it is small in the high-energy regions. Now, it becomes intriguing to investigate the relationship between the increase in temperature and the electric dipole polarizability ($\alpha_D$) through the integrated quantity of the $m_{-1}$ moment. The $\alpha_D$ contains information on the overall E1 transition strength but weights more the low-energy excitations. Furthermore, exploration of the evolution of dipole polarizability with temperature can be pivotal for the EOS of neutron-rich matter at higher temperatures. 

In Figure \ref{fig5}, the $\alpha_D$ values are presented as a function of the mass number for two sets of isotopes, Ca (upper panel) and Sn (lower panel), within the temperature range of $T = 0-2$ MeV. In addition to the increase with the neutron excess, for all isotopes, the dipole polarizability systematically increases with temperature. Due to its strong sensitivity to temperature, especially for neutron-rich nuclei, dipole polarizability can be identified as a reasonable integrated quantity to explore the sensitivity of the dipole nuclear response to finite temperature effects.
In particular, the $\alpha_D$ of neutron-rich nuclei shows a rapid increment with temperature growth, which suggests that temperature effects could play a significant role in neutron-rich matter.

\section{Summary\label{sec:summary}}
A fully self-consistent FT-RQRPA framework has been developed for the study the thermal effects in isovector electric dipole excitations in $^{40-60}$Ca and $^{100-140}$Sn isotopic chains. We employ the relativistic density-dependent point coupling (DD-PCX) interaction to calculate the nuclear properties across temperatures ranging from $T=0$ to $2$ MeV. First, the temperature and isotopic dependence of the low- and high-energy parts of the E1 response are investigated. The E1 strength distributions in the GDR region are slightly modified at the considered temperature range. Furthermore, new low-energy excitations begin to emerge, particularly above the critical temperature ($T_c$), in the neutron-rich Ca and Sn nuclei. These low-energy excitations are observed for $E<$ 5 MeV in both Ca and Sn neutron-rich nuclei, which are mainly formed due to single neutron configurations. These results are also verified by the cumulative sum of low-energy E1 strength that increases with increasing temperature. Also, the centroid energy $E_c$ decreases rapidly with increasing temperature, indicating the dominance of new excitations in the much lower energy region of E1 strength. Additionally, the dipole polarizability $\alpha_D$ is investigated for Ca and Sn isotopic chains, which is significantly altered with the inclusion of temperature and neutron number in the nucleus. It is shown that dipole polarizability represents a reasonable integrated quantity to explore the sensitivity of dipole transition strength to finite temperature, especially for neutron-rich nuclei.

The present study highlights the significance of temperature and pairing effects in dipole excitations in nuclei, together with their isotopic dependence. In the forthcoming study, the FT-RQRPA introduced in this work will be employed in microscopic calculation of $\gamma$-ray strength functions at finite temperature relevant for nuclear reaction studies in hot astrophysical environments. The theory framework established in this work could also be further extended for future studies of the evolution of magnetic dipole transition strength distributions with temperature, which is currently mainly unknown but could also be of relevance for studies of $(n,\gamma)$ reactions involving nuclei at finite temperatures.
\\
\begin{acknowledgments}
This work is supported by the QuantiXLie Centre of Excellence, a project co-financed by the Croatian Government and European Union through the European Regional Development Fund, the Competitiveness and Cohesion Operational Programme (KK.01.1.1.01.0004). E.Y. acknowledges support from the Science and Technology Facilities Council (UK) through grant ST/Y000013/1.
\end{acknowledgments}
\bibliographystyle{apsrev4-2}
\bibliography{E1.bib}
\end{document}